# An Approach to Elastoplasticity at Large Deformations

**K.Y. Volokh**

*Faculty of Civil and Environmental Engineering*

*Technion – Israel Institute of Technology, Haifa 32000, Israel*

*(E-mail: cvolokh@technion.ac.il)*

*In appreciation of R.S. Rivlin (1915-2005)*

## Abstract

While elastoplasticity theories at small deformations are well-established for various materials, elastoplasticity theories at large deformations are still a subject of controversy and lively discussions. Among the approaches to finite elastoplasticity two became especially popular. The first, implemented in the commercial finite element codes, is based on the introduction of a hypoelastic constitutive law and the additive elastic-plastic decomposition of the deformation rate tensor. Unfortunately, the use of hypoelasticity may lead to a nonphysical creation or dissipation of energy in a closed deformation cycle. In order to replace hypoelasticity with hyperelasticity the second popular approach based on the multiplicative elastic-plastic decomposition of the deformation gradient tensor was developed. Unluckily, the latter theory is not perfect as well because it introduces intermediate plastic configurations, which are geometrically incompatible, non-unique, and, consequently, fictitious physically.

In the present work, an attempt is made to combine strengths of the described approaches avoiding their drawbacks. Particularly, a tensor of the plastic deformation rate is introduced in the additive elastic-plastic decomposition of the velocity gradient. This tensor is used in the flow rule defined by the generalized isotropic Reiner-Rivlin fluid. The tensor of the plastic deformation rate is also used in an evolution equation that allows calculating an elastic strain tensor which, in its turn, is used in the hyperelastic constitutive law. Thus, the present approach employs hyperelasticity and the additive decomposition of the velocity gradient avoiding nonphysical hypoelasticity and the multiplicative decomposition of the deformation gradient associated with incompatible plastic configurations. The developed finite elastoplasticity framework for isotropic materials is specified to extend the classical $J_2$-theory of metal plasticity to large deformations and the simple shear deformation is analyzed.

# 1. Introduction

Small deformation elastoplasticity is a well-established theory: Hill (1950); Kachanov (1971); Lubliner (1990); Khan and Huang (1995); Simo and Hughes (1998); Lubarda (2001); Haupt (2002); de Souza Neto et al (2008); Gurtin et al (2010). Unfortunately, the small deformation elastoplasticity is not suitable for some important applications. For example, it is impossible to describe the processes of the metal forming or the large-scale geomechanical flow using small deformations. Also the plasticity problems concerning the structural (Hutchinson, 1974) and material (Needleman and Tvergaard, 1983) instabilities cannot be convincingly posed within the small deformation framework.

Large deformation elastoplasticity is a subject of lasting controversy. A comprehensive review, including 441 references, of elastoplasticity beyond small deformations was done by Xiao et al (2006). Here, we will not follow the historical path of the development of large deformation elastoplasticity but focus on the main ideas.

A direct extension of the small deformation elastoplasticity to the large deformation one involves the additive elastic-plastic decomposition of the deformation rate tensor (Hill, 1958; Hill, 1959; Prager, 1960)

$$\mathbf{D} = \mathbf{D}^e + \mathbf{D}^p . \qquad (1)$$

This decomposition mimics the decomposition of strain rates in the small deformation elastoplasticity. The plastic part of the decomposition is defined by a flow rule analogously to the small deformation elastoplasticity while the elastic part of the decomposition is defined by the hypoelasticity theory proposed by Truesdell (1955). In the general form hypoelasticity can be set as

$$\mathbf{D}^e = \mathbf{C}(\boldsymbol{\sigma}) : \overset{\triangledown}{\boldsymbol{\sigma}} , \qquad (2)$$

where $\mathbf{C}(\boldsymbol{\sigma})$ is a fourth-order tensor of the elastic compliances, which can depend on stresses, and $\overset{\triangledown}{\boldsymbol{\sigma}}$ is an objective rate of the Cauchy stress.

The basic idea behind hypoelasticity is to introduce elasticity in the rate form. It does not work, unfortunately, because the constitutive law (2) can lead to dissipation or creation of energy in a closed deformation cycle. This drawback of hypoelasticity was understood by Truesdell himself (Truesdell and Noll, 1965). Interestingly, besides the problems with the energy conservation the hypoelastic constitutive law can lead to nonphysical stress oscillations in simple shear (Khan and Huang, 1995).

Despite the noted physical shortcomings, hypoelasticity is a popular computational tool: Huespe et al (2011); McMeeking and Rice (1975); Voyiadjis and Kattan (1992). It is usually argued that the shortcomings of hypoelasticity become sound when elastic deformations get large while in the case of small elastic deformations the use of hypoelasticity is safe (Khan and Huang, 1995; Simo and Hughes, 1998; Xiao et al, 2006). In this argument a definition of small elastic deformations might be difficult. It should not be missed also that while elastic strains can be really small (tenth of percent) the elastic rotations can be large as it often occurs in mechanics of thin-walled structures.

Alternatively to the hypoeleasticity-based formulation discussed above and almost simultaneously in time the formulation based on the multiplicative elastic-plastic decomposition of the deformation gradient tensor was developed (Kroner, 1960; Lee and Liu, 1967; Lee, 1969)

$$\mathbf{F} = \mathbf{F}^e \mathbf{F}^p. \qquad (3)$$

According to this decomposition every material point undergoes two successive mappings corresponding to the plastic and elastic deformations. This approach allows using hyperelasticity for a description of elastic deformations. This is a way to get rid of nonphysical hypoelasticity. Nonetheless, the multiplicative decomposition of the deformation gradient is not physically perfect either. The problem is that decomposition (3) introduces stress-relaxed intermediate configurations (after $\mathbf{F}^p$ mapping) in the vicinity of all material points and these configurations are geometrically incompatible. They generally form an abstract mathematical manifold beyond the physical Euclidian space. Even worse, such configurations cannot be defined uniquely and they are isomorphic under superposed rigid rotations. Although the said is probably enough to question the physical aspects of the multiplicative decomposition its formal use in computations might still be reasonable: Arghavani et al (2011); Gurtin (2010); Henann and Anand (2009); Lele and Anand (2009); Thamburaja (2010). The reader is advised to consult Naghdi's (1990) review for the criticism of the theories based on (3). Remarkably, Naghdi's criticism was largely ignored in the subsequent literature.

In view of the drawbacks of the hypoelasticity- and the multiplicative decomposition- based approaches it is worth noticing that another approach was proposed by Green and Naghdi (1965) and developed by Naghdi and his collaborators and followers. In this approach a plastic Green strain $\mathbf{E}^p$ is introduced as a primitive variable and the elastic deformation is described by the difference between the total and plastic Green strains: $\mathbf{E} - \mathbf{E}^p$. The mathematical purity and technical simplicity of this approach are appealing. Its physical basis, however, is arguable on the principal grounds. Indeed, the reference or initial material configuration is the very heart of the Naghdi formulation while materials undergoing plastic flow cannot remember this reference configuration. Only elastic deformations have a perfect memory and the preference of a reference configuration. Flowing materials have no preference to the reference and constitutive equations of the plastic flow should be formulated with respect to the current configuration. Besides, it is desirable that in the presence of plastic flows the elastic deformation should refer to the current material configuration as well.

The fact that the elastic deformations should refer to the current material configuration during plastic flows was realized by Eckart (1948), who introduced inelasticity through a description of the evolving elastic metric. This line of thought was also followed by Leonov (1976); Rubin (2009); Rubin and Ichihara (2010), for example. The approach of the present work is also based on the constitutive description referring to the current material configuration and it further generalizes the ideas of Eckart. Particularly, we relax the restriction on the elastic deformations only and we include the plastic deformations into consideration. Our approach has three ingredients: Rivlin's hyperelasticity; generalized Reiner-Rivlin's non-Newtonian fluidity; and the evolution equation linking elastic strains and plastic deformation rates. Thus, our approach allows describing both elasticity and plasticity at large deformations and it is potentially applicable to a variety of materials ranging from soft polymers to hard metals. We emphasize, however, that we restrict the considerations of the present work by the isotropic material response only.

The paper is organized as follows. The field equations of the classical local continuum mechanics are briefly reviewed in Section 2. The general constitutive framework of the large deformation elastoplasticity is presented in Section 3. This framework is specialized for metals in Section 4 where a large deformation extension of the $J_2$-theory of metal plasticity is presented. The latter theory is used for analysis of simple shear in Section 5. A short summary of the proposed approach is made Section 6.

## 2. Field equations

In continuum mechanics the atomistic or molecular structure of material is approximated by a continuously distributed set of the so-called material points. The continuum material point is an abstraction that is used to designate a small representative volume of real material including many atoms and molecules. A material point that occupies position $\mathbf{x}$ in the reference configuration moves to position $\mathbf{y}(\mathbf{x})$ in the current configuration of the continuum. The deformation in the vicinity of the material point can be completely described by the deformation gradient tensor

$$\mathbf{F} = \frac{\partial \mathbf{y}}{\partial \mathbf{x}} \, . \qquad (4)$$

Introducing the velocity vector as a material time derivative of the current placement of a material point

$$\mathbf{v} = \frac{d\mathbf{y}}{dt} = \dot{\mathbf{y}} \, , \qquad (5)$$

it is possible to describe the time dependent deformation changes with the help of the velocity gradient tensor

$$\mathbf{L} = \frac{\partial \mathbf{v}}{\partial \mathbf{y}} = \dot{\mathbf{F}} \mathbf{F}^{-1} . \tag{6}$$

Neglecting inertia and body forces it is possible to write down the linear and angular momentum balance laws in the following forms accordingly

$$\operatorname{div} \boldsymbol{\sigma} = \mathbf{0} , \tag{7}$$

$$\boldsymbol{\sigma} = \boldsymbol{\sigma}^T , \tag{8}$$

where the divergence operator is calculated with respect to current coordinates $\mathbf{y}$; and $\boldsymbol{\sigma}$ is the Cauchy tensor of true stresses.

The balance of linear momentum on the body surface reads

$$\boldsymbol{\sigma} \mathbf{n} = \bar{\mathbf{t}} \tag{9}$$

where $\bar{\mathbf{t}}$ is a prescribed traction on the surface with the unit outward normal $\mathbf{n}$.

Alternatively to (9) a surface boundary condition can be imposed on placements

$$\mathbf{y} = \bar{\mathbf{y}} , \tag{10}$$

where the barred quantity is prescribed.

Equations (7) and (9) describe equilibrium in the spatial or Eulerian form. It is more convenient sometimes to consider the referential position of material points, $\mathbf{x}$, as an independent variable and reformulate the volumetric, (7), and surface, (9), equilibrium equations in the referential or Lagrangean form

$$\operatorname{Div} \mathbf{P} = \mathbf{0} , \tag{11}$$

$$\mathbf{P} \mathbf{n}_0 = \bar{\mathbf{t}}_0 , \tag{12}$$

where 'Div' operator is with respect to referential coordinates $\mathbf{x}$; $\mathbf{P}$ is the 1$^{\text{st}}$ Piola-Kirchhoff stress tensor; $\mathbf{t}_0$ is traction per unit area of the reference surface with the unit outward normal $\mathbf{n}_0$; and the barred quantity is prescribed.

The Eulerian and Lagrangean quantities are related as follows

$$\mathbf{n} = \mathbf{F}^{-T} \mathbf{n}_0 \left| \mathbf{F}^{-T} \mathbf{n}_0 \right|^{-1} , \tag{13}$$

$$\boldsymbol{\sigma} = J^{-1} \mathbf{P} \mathbf{F}^T , \tag{14}$$

$$\mathbf{t} = \mathbf{t}_0 J^{-1} \left| \mathbf{F}^{-T} \mathbf{n}_0 \right|^{-1} , \tag{15}$$

$$J = \det \mathbf{F} . \tag{16}$$

The filed equations should be completed with the constitutive equations.

## 3. Constitutive equations

Elastoplastic deformations exhibit features of both elasticity and non-Newtonian fluidity so profoundly developed by Rivlin (Barenblatt and Joseph, 1997). In the present section we describe the elasticity and plasticity/fluidity components of the theory separately starting with a rheological model.

### 3.1 Rheological model

The purpose of a rheological model is to create a primitive prototype of a three-dimensional theory. Though different tensorial formulations could be proposed for the same toy prototype they would share similar qualitative features. Specifically, we choose the successively joined spring and friction elements shown in Fig. 1 as a rheological model of elastoplasticity.

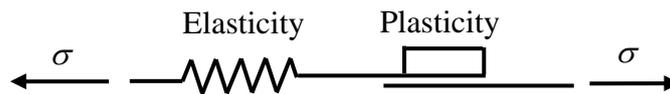

Fig. 1. Rheological model for elastoplacticity at large strains.

Here the spring element is related with elasticity while the friction element is related with plasticity. Remarkably, the friction element can also exhibit a characteristic mechanism of the interlayer friction in liquids, especially, the non-Newtonian ones. Thus, generally, we associate plasticity with non-Newtonian fluidity.

It is crucial for a three-dimensional tensorial formulation, which can stem from the rheological model, that stresses in elasticity and plasticity are equal because the spring and friction elements are joined successively.

### 3.2 Kinematics

Following our discussion in Introduction we remind the reader that various approaches exist to describe deformations of the elements of the rheological model in the case of three-dimensional theory. We choose the following additive elastic-plastic decomposition of the velocity gradient, which is arguably the most appealing physically,

$$\mathbf{L} = \mathbf{L}^e + \mathbf{L}^p . \tag{17}$$

The choice of the velocity gradient for a description of kinematics is natural in the cases of flow. We also notice that the additive decomposition does not introduce the hierarchy of deformations contrary to the multiplicative decomposition.

We further decompose the elastic and plastic parts of the velocity gradient into symmetric and skew-symmetric tensors as follows

$$\mathbf{L}^e = \mathbf{D}^e + \mathbf{W}^e, \quad \mathbf{D}^e = \frac{1}{2}(\mathbf{L}^e + \mathbf{L}^{eT}), \quad \mathbf{W}^e = \frac{1}{2}(\mathbf{L}^e - \mathbf{L}^{eT}), \tag{18}$$

$$\mathbf{L}^p = \mathbf{D}^p + \mathbf{W}^p, \quad \mathbf{D}^p = \frac{1}{2}(\mathbf{L}^p + \mathbf{L}^{pT}), \quad \mathbf{W}^p = \frac{1}{2}(\mathbf{L}^p - \mathbf{L}^{pT}). \tag{19}$$

Here $\mathbf{D}^e$ and $\mathbf{D}^p$ are the elastic and plastic deformation rate tensors accordingly; and $\mathbf{W}^e$ and $\mathbf{W}^p$ are the elastic and plastic spin tensors accordingly.

We assume that the plastic spin is zero

$$\mathbf{W}^p = \mathbf{0}, \tag{20}$$

and, consequently,

$$\mathbf{L} = \mathbf{L}^e + \mathbf{D}^p. \tag{21}$$

Decomposition (21) is a mathematical expression of the separated kinematic response of the spring and friction elements of the rheological model.

We emphasize that assumption (20) corresponds to the isotropic material response. If the response is anisotropic then a constitutive equation for the plastic spin should be defined (Dafalias, 1984; Dafalias, 1985).

### 3.3 Elasticity

The constitutive law describing the elastic behavior of the spring in the rheological model is a generalization of the Rivlin 3D isotropic hyperelastic solid

$$\boldsymbol{\sigma} = 2I_3^{-1/2}(I_3\psi_3\mathbf{1} + (\psi_1 + I_1\psi_2)\mathbf{G} - \psi_2\mathbf{G}^2), \tag{22}$$

where $\psi$ is the elastic strain energy function and $\mathbf{1}$ is the second order identity tensor; invariants are

$$I_1 = \text{tr}\mathbf{G}, \quad I_2 = \{(\text{tr}\mathbf{G})^2 - \text{tr}(\mathbf{G}^2)\}/2, \quad I_3 = \det\mathbf{G}, \tag{23}$$

and

$$\psi_i \equiv \frac{\partial\psi(I_1, I_2, I_3)}{\partial I_i}. \tag{24}$$

Symmetric elastic strain $\mathbf{G} = \mathbf{G}^T$ is *not* generally the left Cauchy-Green tensor used by Rivlin and we postpone its definition to Section 3.5.

### *3.4 Plasticity*

The constitutive law describing the plastic behavior of the friction element in the rheological model – *the flow rule* – is a generalization of the Reiner-Rivlin model for an isotropic fluid

$$\boldsymbol{\sigma} = \beta_1 \mathbf{1} + \beta_2 \mathbf{D}^p + \beta_3 \mathbf{D}^{p\,2}, \qquad (25)$$

where the response functionals depend on the history of the plastic deformation rate and Cauchy stress

$$\beta_1 = \hat{\beta}_1(\boldsymbol{\sigma}, \mathbf{D}^p), \quad \beta_2 = \hat{\beta}_2(\boldsymbol{\sigma}, \mathbf{D}^p), \quad \beta_3 = \hat{\beta}_3(\boldsymbol{\sigma}, \mathbf{D}^p). \qquad (26)$$

We note, in view of (26), that constitutive equation (25) is *history-dependent* and *implicit*.

Equations (25)-(26) present a general description of plasticity while it is argued that an additional yield condition should be obeyed during the plastic deformation

$$f(\boldsymbol{\sigma}, \mathbf{D}^p) = 0. \qquad (27)$$

The physical basis for the yield condition (27) is open for discussion. Amazingly, the yield condition can be a blessing for an analytical solution or it can be a pain in the neck for a numerical procedure. The rate of the yield constraint (27) is often used to derive the so-called plastic multiplier accounting for the history of inelastic deformations. We discuss this issue below concerning the theory of metal plasticity.

### *3.5 Evolution equation*

We remind the reader that symmetric elastic strain $\mathbf{G} = \mathbf{G}^T$ is *not* generally the left Cauchy-Green tensor used by Rivlin. The elastic strain is *defined* as a solution of the evolution equation

$$\dot{\mathbf{G}} - \mathbf{L}^e \mathbf{G} - \mathbf{G} \mathbf{L}^{eT} = \mathbf{0}, \qquad (28)$$

which can be rewritten, accounting for decomposition (21), in the form

$$\dot{\mathbf{G}} - \mathbf{L}\mathbf{G} - \mathbf{G}\mathbf{L}^T + \mathbf{D}^p\mathbf{G} + \mathbf{G}\mathbf{D}^p = \mathbf{0} \qquad (29)$$

with the initial condition

$$\mathbf{G}(t=0) = \mathbf{1}. \qquad (30)$$

It should not be missed that the material response is isotropic otherwise the tensor of the plastic deformation rate should be replaced with the tensor of the plastic deformation gradient in (29): $\mathbf{D}^p \to \mathbf{L}^p = \mathbf{D}^p + \mathbf{W}^p$; where the plastic spin, $\mathbf{W}^p$, should be defined by a constitutive law.

The initial value problem (29)-(30) gives the crucial connection between elastic and plastic deformations. This connection is motivated by the general kinematic identity, which is correct independently of the character of deformation: $\dot{\mathbf{B}} = \mathbf{LB} + \mathbf{BL}^T$; where $\mathbf{B} = \mathbf{FF}^T$ is the left Cauchy-Green tensor. Thus, we assumed that the mentioned kinematic identity should be obeyed for the purely elastic deformations as well.

In the absence of plastic deformations, $\mathbf{D}^p = \mathbf{0}$, equations (29)-(30) reduce to

$$\dot{\mathbf{G}} - \mathbf{LG} - \mathbf{GL}^T = \mathbf{0}, \quad \mathbf{G}(t = 0) = \mathbf{1}, \qquad (31)$$

and the solution of (31) is the left Cauchy-Green tensor

$$\mathbf{G} = \mathbf{B} = \mathbf{FF}^T. \qquad (32)$$

**Remark 1** We notice that the evolution equation (28) is invariant under the superposed rigid body motion. Indeed, let us designate $\mathbf{Q}$ a proper orthogonal tensor describing the superposed rigid body rotation. Then starring the rotated quantities we have

$$\mathbf{G}* = \mathbf{QGQ}^T, \qquad (33)$$

$$\mathbf{L}* = \mathbf{QLQ}^T + \dot{\mathbf{Q}}\mathbf{Q}^T = \mathbf{Q}(\mathbf{L}^e + \mathbf{D}^p)\mathbf{Q}^T + \dot{\mathbf{Q}}\mathbf{Q}^T = \mathbf{L}^e * + \mathbf{D}^p *, \qquad (34)$$

where

$$\mathbf{L}^e * = \mathbf{QL}^e\mathbf{Q}^T + \dot{\mathbf{Q}}\mathbf{Q}^T, \qquad (35)$$

$$\mathbf{D}^p * = \mathbf{QD}^p\mathbf{Q}^T. \qquad (36)$$

By a direct calculation with account of (33)-(36) we have

$$\dot{\mathbf{G}}* - \mathbf{L}^e * \mathbf{G}* - \mathbf{G}*\mathbf{L}^e *^T = \mathbf{Q}(\dot{\mathbf{G}} - \mathbf{L}^e\mathbf{G} - \mathbf{GL}^{eT})\mathbf{Q}^T. \qquad (37)$$

**Remark 2** *In some cases*, it is possible to completely exclude the notion of the plastic deformation from the theory. Indeed, let us assume that the flow rule (25)-(26) is resolved with respect to the plastic deformation rate, $\mathbf{D}^p$. Then, the plastic deformation rate can be expressed as a function of the stress, $\boldsymbol{\sigma}$. The stress, in its turn, is a function of elastic strains, $\mathbf{G}$, through the hyperelastic law (22). Thus, the plastic deformation rate depends on the elastic strain and we can rewrite the evolution law (29) in the form

$$\dot{\mathbf{G}} - \mathbf{LG} - \mathbf{GL}^T + \mathbf{K}(\mathbf{G}) = \mathbf{0} \qquad (38)$$

where $\mathbf{K}(\mathbf{G})$ is a function of the elastic strain.

Examples of the specific choice of $\mathbf{K}(\mathbf{G})$ can be found in: Eckart (1948); Leonov (1976); Rubin and Ichihara (2010). We emphasize, however, that the reduction of the problem to equation (38) is not always possible. We will show below, for example, that the $J_2$-theory of metal plasticity with isotropic hardening cannot be generally reduced to the simple framework presented by (38).

### 3.6 Dissipation

Let us examine the dissipation inequality

$$D = \boldsymbol{\sigma} : \mathbf{D} - I_3^{-1/2} \dot{\psi} \ge 0 . \tag{39}$$

Decomposing the deformation rate and substituting elastic strains the dissipation takes form

$$D = \boldsymbol{\sigma} : \mathbf{D}^e + \boldsymbol{\sigma} : \mathbf{D}^p - I_3^{-1/2} \frac{\partial \psi}{\partial \mathbf{G}} : \dot{\mathbf{G}} \ge 0 . \tag{40}$$

We transform the third term in (40) with account of evolution equation (28) as follows

$$\frac{\partial \psi}{\partial \mathbf{G}} : \dot{\mathbf{G}} = \frac{\partial \psi}{\partial \mathbf{G}} : \mathbf{L}^e \mathbf{G} + \frac{\partial \psi}{\partial \mathbf{G}} : \mathbf{G} \mathbf{L}^{eT} = 2 \frac{\partial \psi}{\partial \mathbf{G}} \mathbf{G} : \mathbf{L}^e = 2 \frac{\partial \psi}{\partial \mathbf{G}} \mathbf{G} : \mathbf{D}^e . \tag{41}$$

Substituting (41) in (40) we get

$$D = (\boldsymbol{\sigma} - 2 I_3^{-1/2} \frac{\partial \psi}{\partial \mathbf{G}} \mathbf{G}) : \mathbf{D}^e + \boldsymbol{\sigma} : \mathbf{D}^p \ge 0 . \tag{42}$$

We notice that the expression in the parentheses equals zero by virtue of the hyperalstic constitutive law

$$\boldsymbol{\sigma} = 2 I_3^{-1/2} \frac{\partial \psi}{\partial \mathbf{G}} \mathbf{G} = 2 I_3^{-1/2} (I_3 \psi_3 \mathbf{1} + (\psi_1 + I_1 \psi_2) \mathbf{G} - \psi_2 \mathbf{G}^2) . \tag{43}$$

Thus, the dissipation inequality reduces to

$$D = \boldsymbol{\sigma} : \mathbf{D}^p \ge 0 . \tag{44}$$

We substitute (25) in (44) as follows

$$D = (\beta_1 \mathbf{1} + \beta_2 \mathbf{D}^p + \beta_3 \mathbf{D}^{p2}) : \mathbf{D}^p \ge 0 . \tag{45}$$

Taking into account that the rate of the plastic deformation is a symmetric tensor, $\mathbf{D}^p = \mathbf{D}^{pT}$, we can rewrite (45) in a more compact form

$$D = \beta_1 \mathrm{tr}(\mathbf{D}^p) + \beta_2 \mathrm{tr}(\mathbf{D}^{p2}) + \beta_3 \mathrm{tr}(\mathbf{D}^{p3}) \ge 0 . \tag{46}$$

In this way, the dissipation inequality imposes a restriction on the plastic response functionals $\beta_1, \beta_2, \beta_3$ and the processes of plastic flow.

## 4. Metal plasticity at large deformations

In this section we cast the classical $J_2$ small deformation elastoplasticity in the general large deformation framework developed in the previous section.

### 4.1 Elasticity

Various hyperelastic isotropic models can be developed, which reduce to the Hooke law at small deformations. We choose here the Ciarlet (1988) proposal for the elastic strain energy function

$$\psi = \frac{\lambda}{4}(I_3 - 1) - \frac{\lambda + 2\mu}{4}\ln I_3 + \frac{\mu}{2}(I_1 - 1), \tag{47}$$

where $\lambda$ and $\mu$ are the Lame constants.

Substituting (47) in (22) we derive the constitutive law for elastic deformations – the spring element of the rheological model – as follows

$$\boldsymbol{\sigma} = I_3^{-1/2}\{\frac{\lambda}{2}(I_3 - 1)\mathbf{1} + \mu(\mathbf{G} - \mathbf{1})\}. \tag{48}$$

### 4.2 Plasticity

In the case of metal plasticity we set the plastic response functions in the following form

$$\beta_1 = \frac{1}{3}\text{tr}\boldsymbol{\sigma}, \quad \beta_2 = \frac{2\tilde{\sigma}}{3\alpha}, \quad \beta_3 = 0, \tag{49}$$

where $\alpha > 0$ is the plastic multiplier and

$$\tilde{\sigma} = \sqrt{\frac{3}{2}\text{dev}\boldsymbol{\sigma} : \text{dev}\boldsymbol{\sigma}} = \sqrt{\frac{3}{2}(\boldsymbol{\sigma} : \boldsymbol{\sigma} - \frac{1}{3}(\text{tr}\boldsymbol{\sigma})^2)} \tag{50}$$

is the von Mises equivalent stress.

Substituting (49) in (25) we get the familiar flow rule

$$\mathbf{D}^p = \frac{3\alpha}{2\tilde{\sigma}}\text{dev}\boldsymbol{\sigma}, \tag{51}$$

where the deformation rate is equal to the strain rate in the case of small deformations.

One of the reasons for the choice of the response functions in the form (49) is the necessity to obey the plastic incompressibility condition

$$\text{tr}\mathbf{D}^p = 0. \tag{52}$$

The yield condition takes form

$$f(\boldsymbol{\sigma}, \tilde{\varepsilon}) = \tilde{\sigma}(\boldsymbol{\sigma}) - \sigma_y(\tilde{\varepsilon}) = 0, \tag{53}$$

$$\tilde{\varepsilon} = \int \dot{\tilde{\varepsilon}}dt, \quad \dot{\tilde{\varepsilon}} = \sqrt{\frac{2}{3}\mathbf{D}^p : \mathbf{D}^p}, \tag{54}$$

where $\sigma_y$ is the yield stress and the accumulated plastic strain, $\tilde{\varepsilon}$, is introduced to account for the isotropic hardening.

Based on the flow rule (51) we can derive the relationship between the rate of the effective plastic strain, $\dot{\tilde{\varepsilon}}$, and the plastic multiplier

$$\underbrace{\mathbf{D}^p : \mathbf{D}^p}_{3\dot{\tilde{\varepsilon}}^2/2} = \left(\frac{3\alpha}{2\tilde{\sigma}}\right)^2 \underbrace{\text{dev}\boldsymbol{\sigma} : \text{dev}\boldsymbol{\sigma}}_{2\tilde{\sigma}^2/3}, \tag{55}$$

and, consequently, we get

$$\dot{\tilde{\varepsilon}} = \alpha \,.$$ (56)

The plastic multiplier is obtained from the following consistency condition

$$\dot{f} = \frac{\partial f}{\partial \boldsymbol{\sigma}} : \dot{\boldsymbol{\sigma}} + \frac{\partial f}{\partial \tilde{\varepsilon}} \dot{\tilde{\varepsilon}} = 0 \,.$$ (57)

We can calculate the stress increment accounting for (29) as follows

$$\dot{\boldsymbol{\sigma}} = \frac{\partial \boldsymbol{\sigma}}{\partial \mathbf{G}} : \dot{\mathbf{G}} = \frac{\partial \boldsymbol{\sigma}}{\partial \mathbf{G}} : (\mathbf{LG} + \mathbf{GL}^T - \mathbf{D}^p \mathbf{G} - \mathbf{GD}^p) \,.$$ (58)

Substituting (51) in (58) we get

$$\dot{\boldsymbol{\sigma}} = \frac{\partial \boldsymbol{\sigma}}{\partial \mathbf{G}} : \dot{\mathbf{G}} = \frac{\partial \boldsymbol{\sigma}}{\partial \mathbf{G}} : (\mathbf{LG} + \mathbf{GL}^T - \frac{3\alpha}{2\tilde{\sigma}} [(\mathrm{dev}\boldsymbol{\sigma})\mathbf{G} + \mathbf{G}\,\mathrm{dev}\boldsymbol{\sigma}]) \,.$$ (59)

Finally, substituting (56) and (59) in (57) we find the plastic multiplier

$$\alpha = \frac{\dfrac{\partial f}{\partial \boldsymbol{\sigma}} : \dfrac{\partial \boldsymbol{\sigma}}{\partial \mathbf{G}} : (\mathbf{LG} + \mathbf{GL}^T)}{\dfrac{3}{2\tilde{\sigma}} \dfrac{\partial f}{\partial \boldsymbol{\sigma}} : \dfrac{\partial \boldsymbol{\sigma}}{\partial \mathbf{G}} : [(\mathrm{dev}\boldsymbol{\sigma})\mathbf{G} + \mathbf{G}\,\mathrm{dev}\boldsymbol{\sigma}] - \dfrac{\partial f}{\partial \tilde{\varepsilon}}} \,,$$ (60)

where

$$\frac{\partial f}{\partial \boldsymbol{\sigma}} = \frac{\partial \tilde{\sigma}}{\partial \boldsymbol{\sigma}} = \frac{3\mathrm{dev}\boldsymbol{\sigma}}{2\tilde{\sigma}} \,,$$ (61)

and

$$\frac{\partial \boldsymbol{\sigma}}{\partial \mathbf{G}} = I_3^{-1/2} (\frac{\lambda(I_3 + 1) + 2\mu}{4} \mathbf{A}_1 - \frac{\mu}{2} \mathbf{A}_2 + \mu\overline{\mathbf{1}}) \,,$$ (62)

$$\mathbf{A}_1 = \frac{1}{2} \{\mathbf{G}^{-1} \otimes \mathbf{1} + \mathbf{1} \otimes \mathbf{G}^{-1}\} \,,$$ (63)

$$\mathbf{A}_2 = \frac{1}{2} \{\mathbf{G}^{-1} \otimes \mathbf{G} + \mathbf{G} \otimes \mathbf{G}^{-1}\} \,,$$ (64)

$$(\overline{\mathbf{1}})_{mnij} = \frac{1}{2} (\delta_{mi}\delta_{nj} + \delta_{ni}\delta_{mj}) \,.$$ (65)

**Remark 3** We notice that the metal plasticity theory described above cannot be cast in the framework described by equation (38) in Remark 2 because the plastic modulus

$$H = -\frac{\partial f}{\partial \tilde{\varepsilon}} = \frac{\partial \sigma_y}{\partial \tilde{\varepsilon}}$$ (66)

generally depends on the accumulated plastic strain and we cannot get rid of it.

*4.4 Dissipation*

To check the dissipation inequality we substitute (49) and (52) in (46) as follows

$$D = \frac{2\tilde{\sigma}}{3\alpha}\operatorname{tr}(\mathbf{D}^{p2}) \geq 0 .$$ (67)

This inequality is evidently obeyed in the case of plastic deformations because all cofactors are positive.

## 5. Simple shear

We consider the problem of simple shear to illustrate the theory developed in the previous section. This problem allows analytically tracking the machinery of the theory and obtaining a compact final evolution equation, which describes the dependence of the shear stress on the amount of shear.

### 5.1 Kinematics

We start with the deformation law for simple shear

$$\mathbf{y} = \mathbf{x} + \gamma(t)x_2\mathbf{e}_1 ,$$ (68)

where $\mathbf{x}$ and $\mathbf{y}$ are the referential and current positions of a material point accordingly; $\gamma$ is the amount of shear; and $\mathbf{e}_1$ is a base vector.

We further assume that the amount of shear changes in a steady mode with the constant velocity

$$\gamma = \dot{\gamma}t, \quad \dot{\gamma} = \text{constant} .$$ (69)

Based on (68)-(69) we can calculate the velocity vector and the velocity gradient tensor

$$\dot{\mathbf{y}} = \dot{\gamma}x_2\mathbf{e}_1 = \dot{\gamma}y_2\mathbf{e}_1 ,$$ (70)

$$\mathbf{L} = \dot{\gamma}\mathbf{e}_1 \otimes \mathbf{e}_2 .$$ (71)

### 5.2 Elasticity

Before plastic deformations occur we have a purely elastic deformation with the strain tensor equal the left Cauchy-Green tensor

$$\mathbf{G} = \mathbf{B} = \mathbf{F}\mathbf{F}^T = \mathbf{1} + \gamma^2\mathbf{e}_1 \otimes \mathbf{e}_1 + \gamma(\mathbf{e}_1 \otimes \mathbf{e}_2 + \mathbf{e}_2 \otimes \mathbf{e}_1) ,$$ (72)

$$I_3 = \det\mathbf{G} = 1 .$$ (73)

Then, the Cauchy stress tensor takes form

$$\boldsymbol{\sigma} = I_3^{-1/2}\{\frac{\lambda}{2}(I_3 - 1)\mathbf{1} + \mu(\mathbf{G} - \mathbf{1})\} = \sigma_{11}\mathbf{e}_1 \otimes \mathbf{e}_1 + \sigma_{12}(\mathbf{e}_1 \otimes \mathbf{e}_2 + \mathbf{e}_2 \otimes \mathbf{e}_1) ,$$ (74)

$$\sigma_{11} = \mu\gamma^2, \quad \sigma_{12} = \mu\gamma .$$ (75)

We notice that the normal stress and strain appear in (74) and (72) typically of large elastic deformations. The fact that it is necessary to apply the normal stress to maintain simple shear is known by the name of the Poynting effect (Truesdell and Noll, 1965; Barenblatt and Joseph, 1997).

When plastic deformations occur we assume that elastic deformations are small and, consequently, the tensor of elastic strains can be written as follows

$$\mathbf{G} = \mathbf{1} + b(\mathbf{e}_1 \otimes \mathbf{e}_2 + \mathbf{e}_2 \otimes \mathbf{e}_1) , \tag{76}$$

$$I_3 = \det \mathbf{G} = 1 - b^2 \approx 1 . \tag{77}$$

The stress tensor triggered by (76)-(77) takes form

$$\boldsymbol{\sigma} = \sigma_{12}(\mathbf{e}_1 \otimes \mathbf{e}_2 + \mathbf{e}_2 \otimes \mathbf{e}_1) = \mu b(\mathbf{e}_1 \otimes \mathbf{e}_2 + \mathbf{e}_2 \otimes \mathbf{e}_1) . \tag{78}$$

### 5.3 Plasticity

The plasticity description starts with the calculation of the von Mises stress based on (78)

$$\tilde{\sigma} = \sqrt{3\sigma_{12}^2} = \mu\sqrt{3}b . \tag{79}$$

Then, we assume that the isotropic hardening of the yield stress is described by the Ramberg-Osgood formula

$$\sigma_y = E\tilde{\varepsilon}_0 \left( \frac{\tilde{\varepsilon}}{\tilde{\varepsilon}_0} \right)^{1/n} , \tag{80}$$

where $E$ is the material Young modulus; $n$ is a constant of hardening; and $\tilde{\varepsilon}_0$ is a material constant designating the effective strain of the onset of the plastic deformation.

Substituting (79)-(80) in (53) we obtain the yield condition

$$f = \mu\sqrt{3}b - E\tilde{\varepsilon}_0 \left( \frac{\tilde{\varepsilon}}{\tilde{\varepsilon}_0} \right)^{1/n} = 0 , \tag{81}$$

where the Young and shear moduli are related through the Poisson ration: $E/\mu = 2(1+\nu)$ .

We turn to the plastic deformation rate, which can be written as follows,

$$\mathbf{D}^p = \frac{d}{2}(\mathbf{e}_1 \otimes \mathbf{e}_2 + \mathbf{e}_2 \otimes \mathbf{e}_1) , \tag{82}$$

where $d$ is the unknown constant rate of the plastic deformation.

Substituting (82) in (54) we obtain

$$\dot{\tilde{\varepsilon}} = d/\sqrt{3} , \tag{83}$$

and

$$\tilde{\varepsilon} = \int_0^{\gamma/\dot{\gamma}} \dot{\tilde{\varepsilon}} dt = d/\sqrt{3} \int_0^{\gamma/\dot{\gamma}} dt = \frac{\gamma d}{\dot{\gamma}\sqrt{3}} . \tag{84}$$

Substituting (84) in (81) we obtain the yield condition in the form

$$f = \mu\sqrt{3}b - E\tilde{\varepsilon}_0\left(\frac{\gamma d}{\dot{\gamma}\tilde{\varepsilon}_0\sqrt{3}}\right)^{1/n} = 0 , \tag{85}$$

and, consequently, we can find the analytical relationship between unknowns $d$ and $b$

$$d = \frac{\dot{\gamma}\tilde{\varepsilon}_0\sqrt{3}}{\gamma}\left(\frac{\mu\sqrt{3}b}{E\tilde{\varepsilon}_0}\right)^n . \tag{86}$$

### 5.4 Evolution equation

It remains to find unknown $b$ using the evolution equation. Since the approximation has been made concerning the form of $\mathbf{G}$ we will only consider one evolution equation for the shear strain $G_{12}$. In this case (29)-(30) reduce to

$$\dot{b} + d = \dot{\gamma}, \quad b(t=0) = 0 , \tag{87}$$

or, substituting from (86), we have

$$\dot{b} + \frac{\dot{\gamma}\tilde{\varepsilon}_0\sqrt{3}}{\gamma}\left(\frac{\mu\sqrt{3}b}{E\tilde{\varepsilon}_0}\right)^n = \dot{\gamma}, \quad b(t=0) = 0 , \tag{88}$$

Pre-multiplying (88) by shear modulus $\mu$ we obtain the evolution equation for the shear stress during the plastic deformation

$$\dot{\sigma}_{12} + \frac{\mu\dot{\gamma}\tilde{\varepsilon}_0\sqrt{3}}{\gamma}\left(\frac{\sqrt{3}\sigma_{12}}{E\tilde{\varepsilon}_0}\right)^n = \mu\dot{\gamma}, \quad \sigma_{12}(t=0) = 0 , \tag{90}$$

In (90) we can consider the dependence on the amount of shear instead of time. Based on (69) we have

$$\frac{d}{dt} = \dot{\gamma}\frac{d}{d\gamma} , \tag{91}$$

and, substituting (91) in (90),

$$\frac{d\sigma_{12}}{d\gamma} + \frac{\mu\tilde{\varepsilon}_0\sqrt{3}}{\gamma}\left(\frac{\sqrt{3}\sigma_{12}}{E\tilde{\varepsilon}_0}\right)^n = \mu, \quad \sigma_{12}(\gamma=0) = 0 . \tag{92}$$

It remains only to introduce the dimensionless shear stress

$$\bar{\sigma}_{12} = \frac{\sigma_{12}}{\mu} . \tag{93}$$

Substituting (93) in (92) we have

$$\frac{d\bar{\sigma}_{12}}{d\gamma} + \frac{\sqrt{3}\tilde{\varepsilon}_0}{\gamma}\left(\frac{\sqrt{3}\bar{\sigma}_{12}}{2(1+\nu)\tilde{\varepsilon}_0}\right)^n = 1, \quad \bar{\sigma}_{12}(\gamma=0) = 0 , \tag{94}$$

where $\nu$ is the Poisson ratio.

The reader should not forget that the second term on the left hand side of (94) is present during the plastic deformation only. In the case of the purely elastic deformation the second term is omitted and we have the elastic solution: $\sigma_{12} = \mu\gamma$.

Initial-value problem (94) is easily integrated numerically for constants: $\tilde{\varepsilon}_0 = 0.002$; $n = 5$; $\nu = 0.3$. The numerically generated stress-strain curve is shown in Fig. 2.

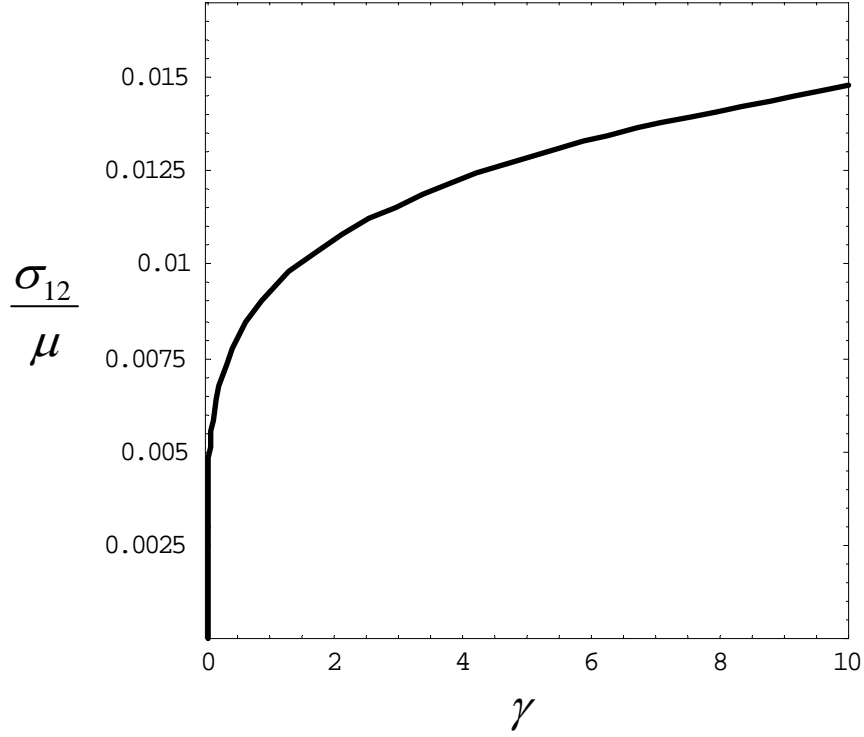

Fig. 2. Stress versus shear in the simple shear problem.

### 5.5 Comparison with the multiplicative decomposition of the deformation gradient

In this section we compare the approach of the present work to the approach based on the multiplicative decomposition of the deformation gradient. We start with some general observations.

First, the additive decomposition of the velocity gradient, which is triggered by the multiplicative decomposition (3), takes the following form

$$\mathbf{L} = \dot{\mathbf{F}}\mathbf{F}^{-1} = \mathbf{L}^e + \mathbf{L}^p, \tag{95}$$

$$\mathbf{L}^e = \dot{\mathbf{F}}^e\mathbf{F}^{e-1}, \quad \mathbf{L}^p = \mathbf{F}^e\dot{\mathbf{F}}^p\mathbf{F}^{p-1}\mathbf{F}^{e-1}, \tag{96}$$

where $\mathbf{L}^e$ and $\mathbf{L}^p$ are the elastic and plastic parts of the velocity gradient accordingly.

Second, the elastic and plastic strains can be written as follows

$$\mathbf{B}^e = \mathbf{F}^e\mathbf{F}^{eT} = \mathbf{F}\mathbf{C}^{p-1}\mathbf{F}^T, \tag{97}$$

$$\mathbf{C}^p = \mathbf{F}^{pT}\mathbf{F}^p. \qquad (98)$$

Third, differentiating (97) with respect to time we get the evolution equation

$$\dot{\mathbf{B}}^e - \mathbf{L}^e\mathbf{B}^e - \mathbf{B}^e\mathbf{L}^{eT} = \mathbf{0}, \qquad (99)$$

which, after some algebraic manipulations and the use of identity $\dot{\mathbf{F}}^{p-1} = -\mathbf{F}^{p-1}\dot{\mathbf{F}}^p\mathbf{F}^{p-1}$, can be rewritten in the following form

$$\dot{\mathbf{B}}^e - \mathbf{L}\mathbf{B}^e - \mathbf{B}^e\mathbf{L}^T = \mathbf{F}\dot{\mathbf{C}}^{p-1}\mathbf{F}^T. \qquad (100)$$

Equation (99) formally coincides with the evolution equation (28) where $\mathbf{G} = \mathbf{B}^e$. However, there is a significant informal difference between them because all quantities in (28) are *independent* while all quantities in (99) *depend* on the elastic, $\mathbf{F}^e$, and plastic, $\mathbf{F}^p$, parts of the deformation gradient.

For example, within the multiplicative decomposition framework one cannot prescribe $\mathbf{F}^e, \mathbf{F}^p$ and $\mathbf{L}^e, \mathbf{L}^p$ independently. The latter means, particularly, that in the case of the plastic isotropy it is impossible to impose conditions of the zero plastic spin on the anti-symmetric part of the plastic velocity gradient

$$\mathbf{W}^p = \frac{1}{2}(\mathbf{L}^p - \mathbf{L}^{pT}) \neq \mathbf{0}. \qquad (101)$$

Thus, even isotropic and isotropically deforming material will produce the plastic spin!

We illustrate this notion on the simple shear deformation considered above. In this case we have for the elastic deformations

$$\mathbf{F}^e = \mathbf{1} + \gamma^e\mathbf{e}_1 \otimes \mathbf{e}_2, \qquad (102)$$

$$(\mathbf{F}^e)^{-1} = \mathbf{1} - \gamma^e\mathbf{e}_1 \otimes \mathbf{e}_2, \qquad (103)$$

$$J^e = \det\mathbf{F}^e \approx 1, \qquad (104)$$

$$\mathbf{L}^e \equiv \dot{\mathbf{F}}^e(\mathbf{F}^e)^{-1} \approx \dot{\gamma}^e\mathbf{e}_1 \otimes \mathbf{e}_2, \qquad (105)$$

where $\gamma^e$ is the amount of elastic shear.

For the plastic deformations we have

$$\mathbf{F}^p = \mathbf{1} + \gamma^p\mathbf{e}_1 \otimes \mathbf{e}_2, \qquad (106)$$

$$(\mathbf{F}^p)^{-1} = \mathbf{1} - \gamma^p\mathbf{e}_1 \otimes \mathbf{e}_2, \qquad (107)$$

$$\mathbf{L}^p \equiv \mathbf{F}^e\dot{\mathbf{F}}^p(\mathbf{F}^p)^{-1}(\mathbf{F}^e)^{-1} \approx \dot{\gamma}^p\mathbf{e}_1 \otimes \mathbf{e}_2, \qquad (108)$$

where $\gamma^p$ is the amount of plastic shear.

The reduced equation (95) follows from (71), (105), and (108)

$$\dot{\gamma} = \dot{\gamma}^e + \dot{\gamma}^p. \qquad (109)$$

Designating

$$\gamma^e = b, \quad \dot{\gamma}^p = d, \tag{110}$$

we obtain the complete set of equations considered in the previous sections.

At this point, one might conclude that the formulations and results of the previous and present sections coincide. Such a conclusion would not be accurate because the approach of the present work predicts zero plastic spin while the multiplicative decomposition approach predicts the following nonzero plastic spin

$$\mathbf{W}^p = \frac{1}{2}(\mathbf{L}^p - \mathbf{L}^{pT}) = \frac{\dot{\gamma}^p}{2}(\mathbf{e}_1 \otimes \mathbf{e}_2 - \mathbf{e}_2 \otimes \mathbf{e}_1). \tag{111}$$

Thus, in the theory developed in the present work the plastic spin is an independent variable and in the case of the isotropic response the plastic spin can be assumed zero and the transition can be done from (28) to (29). The latter transition is impossible within the framework of the multiplicative decomposition of the deformation gradient, where the plastic spin is a dependent variable.

## 6. Concluding remarks

A general constitutive framework for the large deformation isotropic elastoplasticity summarized in the box below was developed in the present study.

$$\boldsymbol{\sigma} = 2I_3^{-1/2}(I_3\psi_3\mathbf{1} + (\psi_1 + I_1\psi_2)\mathbf{G} - \psi_2\mathbf{G}^2),$$

$$\boldsymbol{\sigma} = \hat{\beta}_1(\boldsymbol{\sigma}, \mathbf{D}^p)\mathbf{1} + \hat{\beta}_2(\boldsymbol{\sigma}, \mathbf{D}^p)\mathbf{D}^p + \hat{\beta}_3(\boldsymbol{\sigma}, \mathbf{D}^p)\mathbf{D}^{p2},$$

$$\dot{\mathbf{G}} - \mathbf{LG} - \mathbf{GL}^T + \mathbf{D}^p\mathbf{G} + \mathbf{GD}^p = \mathbf{0}, \quad \mathbf{G}(t=0) = \mathbf{1}.$$

The elastic part of the elastoplastic deformation is described by the elastic strain tensor $\mathbf{G}$, which defines the hyperelastic constitutive law in the first row of the box. The plastic part of the elastoplastic deformation is described by the plastic deformation rate tensor $\mathbf{D}^p$, which defines the generalized non-Newtonian flow rule in the second row of the box. The elastic strain tensor $\mathbf{G}$ and the plastic deformation rate tensor $\mathbf{D}^p$ are related through the evolution equation presented in the third row of the box.

The introduction of the elastic strain tensor $\mathbf{G}$ and hyperelasticity allows avoiding the use of the nonphysical hypoelasticity on the one hand. The introduction of the plastic deformation rate tensor $\mathbf{D}^p$ allows avoiding the use of the multiplicative decomposition of the deformation gradient having a vague physical meaning on the other hand. Thus, the proposed framework avoids the shortcomings of the traditional approaches. Moreover, all constitutive equations and unknowns in the box are referred to the current material configuration, which is physically appealing. Indeed, during plastic deformations a material loses its memory of the initial or reference configuration and it is reasonable to exclude an explicit notion of this configuration from the constitutive formulation.

The proposed framework can be specialized for various materials. As an example of such a specialization the classical small deformation $J_2$-theory of metal plasticity was extended to large deformations in the present work. The analytically tractable example of the simple shear deformation was analyzed. Further examination and use of the proposed general framework will require computer simulations and numerical schemes for the constitutive updating. The latter topic, however, is beyond the scope of the present work.

## Acknowledgement

This work was supported by the General Research Fund at the Technion.